\documentclass{jfm}

\usepackage{graphicx}
 \usepackage{natbib}
\usepackage{color}
\ifCUPmtlplainloaded \else
  \checkfont{eurm10}
  \iffontfound
    \IfFileExists{upmath.sty}
      {\typeout{^^JFound AMS Euler Roman fonts on the system,
                   using the 'upmath' package.^^J}%
       \usepackage{upmath}}
      {\typeout{^^JFound AMS Euler Roman fonts on the system, but you
                   dont seem to have the}%
       \typeout{'upmath' package installed. JFM.cls can take advantage
                 of these fonts,^^Jif you use 'upmath' package.^^J}%
      }
  \else
  \fi
\fi

\ifCUPmtlplainloaded \else
  \checkfont{msam10}
  \iffontfound
    \IfFileExists{amssymb.sty}
      {\typeout{^^JFound AMS Symbol fonts on the system, using the
                'amssymb' package.^^J}%
       \usepackage{amssymb}%

      }{}
  \fi
\fi

\ifCUPmtlplainloaded \else
  \IfFileExists{amsbsy.sty}
    {\typeout{^^JFound the 'amsbsy' package on the system, using it.^^J}%
     \usepackage{amsbsy}}
    {}
\fi

\newcommand\Rey{\mbox{\textit{Re}}}  
\newcommand\Web{\mbox{\textit{We}}}  

\newsavebox{\astrutbox}
\sbox{\astrutbox}{\rule[-5pt]{0pt}{20pt}}

\newcommand\solid[1]{\protect\rule[0.6ex]{#1}{0.1mm}}
\newcommand\dashed[2]{\solid{#1}\hspace{#2}\solid{#1}\hspace{#2}\solid{#1}}

\title[Growth and instability of the liquid rim]{Growth and instability of the liquid rim in the crown splash regime}

\author[G. Agbaglah and R. D. Deegan]%
{G.\ls Agbaglah\thanks{Email address for correspondence: agbagla@umich.edu}, R.\ls D.\ls Deegan\thanks{Email address for correspondence: rddeegan@umich.edu}}

\affiliation{Department of Physics \& Center for the Study of Complex Systems, University of Michigan, Ann Arbor, MI 48109 USA\\}
\begin{document}

\maketitle

\begin{abstract}
We study the formation, growth, and disintegration of jets following impact of a drop on a thin film of the same liquid for $\Web<1000$ and $\Rey<2000$ using a combination of  numerical 
simulations and linear stability theory~\citep{Agbaglah13}.  Our simulations faithfully capture 
this phenomena and are in good agreement with experimental profiles obtained from high-speed X-ray imaging.  
We obtain scaling relations from our simulations and use these 
 as inputs to our stability analysis. The resulting prediction for the most unstable wavelength are in excellent agreement with experimental data.  
Our calculations show that the dominant destabilizing mechanism is a competition between capillarity and inertia but that deceleration 
of the rim provides an additional boost to growth.  We also predict over the entire parameter range of our study the number and timescale for formation of secondary droplets formed during a splash, based on the assumption that the  most unstable mode sets the droplet number.
\end{abstract}


\section{Introduction}
The impact of a drop on a film of the same fluid is ubiquitous in nature and arises in many different contexts such as rain and surf interactions with the air-sea interface, fuel injection systems, spray painting, and atomization. Depending on the governing dimensionless parameters -- mainly the ratio of the film thickness to the drop diameter, and the Reynolds and Weber numbers -- the impact may cause a splash~\citep[see e.g.][]{cossali97,rioboo03,deegan08} herein defined as the generation of secondary droplets. Splashing encompasses a broad variety of qualitatively different morphologies distinguished by their regularity and size-distribution of secondary droplets.  Whether these differences arise from a single mechanism or multiple mechanisms remains an open question. Indeed, the cause of splashing has been revisited often in century since the pioneering studies of \citet{Worthington1879} and has been answered in  many,
seemingly contradictory ways: \citet{Rieber99}, \citet{Bremond06}, and \citet{Zhang:2010} argue for the Rayleigh-Plateau mechanism, \citet{Gueyffier98} 
for a Richtmyer-Meshkov mechanism, \citet{Krechetnikov09a} for a combination of the Richtmyer-Meshkov and Rayleigh-Taylor mechanism,  
\citet{Krechetnikov10} for a combination of the Rayleigh-Plateau and Rayleigh-Taylor mechanism, and \citet{Yarin95} for a nonlinear amplification mechanism 
governed by the eikonal equation. More broadly, Roisman and collaborators~\citep{roisman06,roisman07,roisman10} investigated the linear stability of a receding liquid sheet and concluded that
the liquid rim is subject to a Rayleigh-Plateau and Rayleigh-Taylor instability; a followup study by \citet{Agbaglah13}
that included the growth of the liquid rim in the linear stability analysis concluded that the rim is susceptible to both the Rayleigh-Taylor and Rayleigh-Plateau instabilities but at different times during its evolution: at early times when the rim is decelerating sharply the former dominates and at later times the latter dominates.

In the \emph{crown splash} regime -- the focus of this paper -- the splashing proceeds through a sequence of clearly distinguishable steps: a sheet-like jet emerges and grows outward from the neck of fluid connecting the drop and the pool; the jet's leading edge
is pulled back into the sheet by surface tension and forms a rim as it moves back into the sheet; lastly, the rim develops transverse corrugations that grow, sharpen into fingers, and ultimately pinch-off into secondary droplets.  Because of its high degree of regularity the crown splash offers the simplest scenario in which to examine extant issues on the origin of secondary droplets.

Below we present the results of our numerical study of the initial axisymmetric phase of a crown splash.  The validity of these simulations is demonstrated by their faithful reproduction of the finest details resolvable in experiments.  Using the characteristics of the jet obtained from simulation as inputs for the linear stability theory of \citet{Agbaglah13} we are able to reproduce the experimental results of \citet{Zhang:2010} on the breakup of the crown.  We extend our predictions to regimes for which there are currently no experiments by
extracting from simulations  scaling relationships for the rim's characteristics as a function of Weber and Reynolds number and using these to predict the most unstable wavelength throughout the \emph{crown splash} regime.
We show that the Rayleigh-Plateau mechanism is the dominant mechanism for the parameter range of our study though its growth is significantly enhanced by the Raleigh-Taylor mechanism.

The manuscript is divided into a numerical section and a modeling section.  In the numerical section we describe our numerical technique; validate the results of the simulations with high speed X-ray images; we compute the thickness of the jet and the radius and position of the rim for a variety of Reynolds and Weber numbers and develop a scaling relationship for the jet thickness at the moment it emerges from the impact.  In the modeling section we use these scaling relationship to predict the number of secondary droplets as a function of Weber and Reynolds number.  

 \section{Numerical computations}
\subsection{Numerical method}
We simulate two incompressible fluids, a liquid and a gas, with constant densities $\rho_L$ and $\rho_G$ and constant viscosities $\mu_L$ and $\mu_G$, with the two-fluid Navier-Stokes equations:

\begin{eqnarray*}
 &\rho (\partial_t \textbf{u}+\textbf{u} \cdot {\nabla}
   \textbf{u}) =  - {\nabla} p +  \mu
   \Delta  \textbf{u}+ \sigma \kappa \delta_s \textbf{n},& \\
 &\partial_t \rho + {\nabla} \cdot (\rho \textbf{u}) = 0,&  \\
 &{\nabla} \cdot \textbf{u}= 0. &
\end{eqnarray*}
In this formulation the density $\rho$ and viscosity $\mu$ are constant within each phase and discontinuous at the interface.
The Dirac
distribution function $\delta_s$ expresses the fact that the surface tension
term is concentrated at the interface, $\kappa$ and $\textbf{n}$ being the curvature and the normal of the
interface respectively and $\sigma$ the interfacial tension.

Numerical simulations were performed using GERRIS~\citep{Gerris}, an open-source code where the interface is tracked using a
Volume-of-Fluid method on an octree structured grid allowing adaptive mesh refinement, and the incompressibility condition is
satisfied using a multigrid solver~\citep[see][]{Pop03,GerrisVOF}. This numerical code has been validated with numerous examples and
 used successfully for many different multiphase problems~\citep{fuster:2009a,Gerris-CRAS} including splashing~\citep[see e.g.][]{Thoraval:2012a}.

\subsection{Numerical simulation}
We simulated a spherical liquid drop of diameter $D$ and velocity $U$ impacting normally on a uniform layer of thickness $H$ of the 
same liquid in an initially quiescent gas. The results depend on five parameters: the viscosity and density ratios ($\mu_L/\mu_G$ and 
$\rho_L/\rho_G$), the dimensionless fluid depth $h=H/D$, the liquid Reynolds number
$\Rey=\rho_L {\rm U} {\rm D}/\mu_L$ and  Weber number ${\Web}=\rho_L {\rm U}^2 {\rm D}/\sigma$. 
 We used the axisymmetric formulation of the Navier-Stokes equations described above. Typically we stopped our simulation at 
$t=D/U$ because experiments~(e.g. \citet{Zhang:2010}) show that axial symmetry breaks before this point and hence our axisymmetric simulation 
cannot follow the true dynamics. Our computational domain extends $10D$ in the radial direction and $10D$ in the vertical direction. 
We define $t=0$ as the moment when the drop first touches
the pool; this time is slightly later ( $\simeq 14\ \mu$s) than the usual experimental choice of $t=0$ 
as the moment when the bottom of the drop crosses the undisturbed level of the pool.

\begin{figure}
\centerline{
          \includegraphics[width=0.8\textwidth]{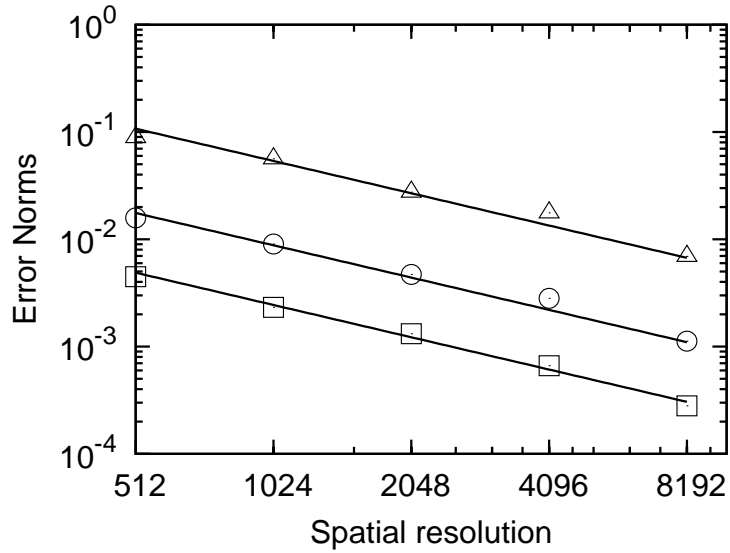}
         }
\caption{Convergence of the numerical simulations for $\Rey=2042$, $\Web=292$, $ h=0.2$, $\mu_L/\mu_G=69$, $\rho_L/\rho_G=709$ parametrized by the $L^1$~($\square$), $L^2$~($\bigcirc$) and $L^{\infty}$~($\bigtriangleup$)  norms on a log-log scale.  The lines correspond to power law fits that yield 
 $2.5/x$, $9/x$ and $55/x$.}
\label{fig:converge}
\end{figure}

We assessed the convergence of the code at 
$\Rey=2042$, $\Web=292$, $\mu_L/\mu_G=69$, $\rho_L/\rho_G=709$  for various dimensionless pool depths in the range 0.2 to 5 by observing the velocity norm as a function of grid size.
We used the $L^1$, $L^2$ and $L^{\infty}$ norms (respectively the average of the absolute values, the root-mean-square norm and the 
maximum absolute value) of the difference between the velocity at a given grid size ($v(X_j, t_k)$) and the velocity at 
the finest grid size ($v_{r}(X_j,t_k)$), at 66 spatial points dispersed over the computational domain and at 20 different times.
The average in time of the computed error are shown in Fig.~\ref{fig:converge}. The convergence is at the first order in the spatial 
resolution.

In the following, a $2^{13}\times 2^{13}$ mesh is used in all the simulations. This corresponds to an initial mesh size of
$\Delta x={ D}/1638$, and it is also the mesh size at the maximum refinement. The mesh is adapted based on three criteria: 1) distance to the interface, 2) curvature of the interface, 3) vorticity magnitude.

\subsection{Validation}
\begin{figure}
\begin{center}

         \includegraphics[width=\textwidth]{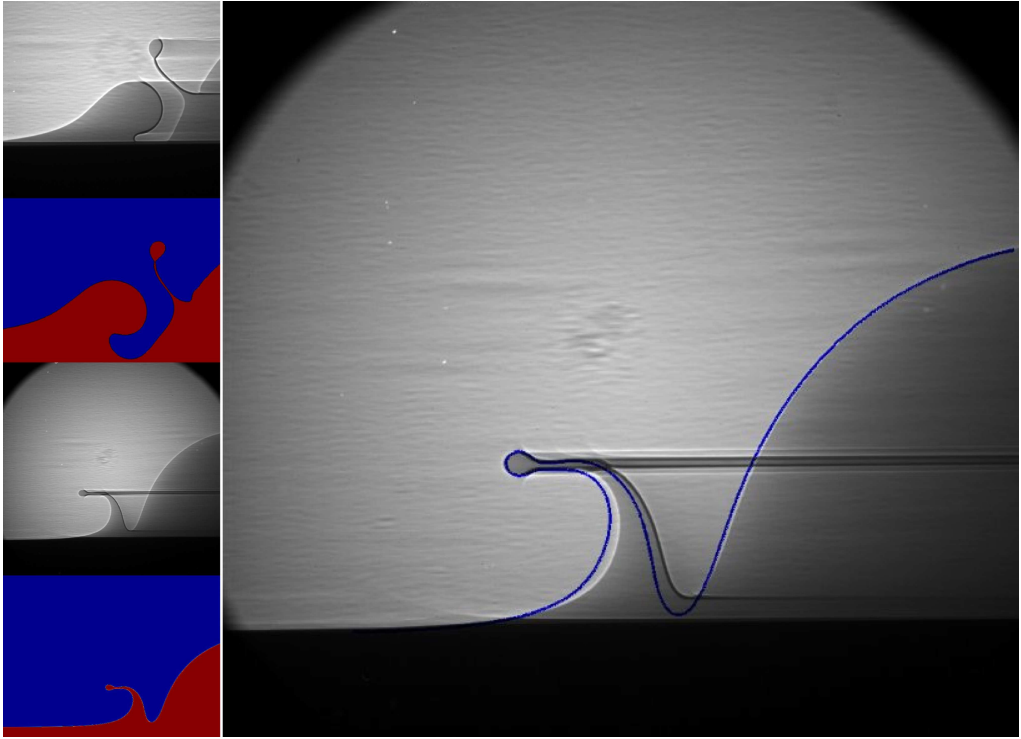}
\end{center}
\caption{Validation. \emph{Left:} Comparison of experimental profiles obtained using high speed X-ray imaging and numerical simulations  for $h=5$, $\Web=451$ and $\Rey=710$ (top),
and  $h=0.2$, $\Web=324$ and $\Rey=2191$ (bottom). \emph{Right:} Superimposed simulation (blue line) and experimental profiles at t=335 $\mu$s ($tU/D=0.468$) for $h=0.2$, $\Web=324$, $\Rey=2191$.}
\label{fig:compar_snap}
\end{figure}

Figure \ref{fig:compar_snap} compares simulation results with the equivalent experiments obtained by using high speed X-ray imaging as described in~\citet{Zhang:2011}.  The simulations at  $h=5$ correspond to the \emph{two-jets}
regime near the transition point from the \emph{one-jet} regime observed experimentally~\citep{Zhang:2011}. Due to the slenderness of the structures (e.g. the ejecta or the gap between the ejecta and the lamella), at these parameter values insufficient grid resolution is readily apparent and thus our comparison of with experiments is a stringent test of our simulation's accuracy.  We find excellent qualitative agreement between simulations and experiments.

As shown in the merged simulation-experiment panel of Fig.~\ref{fig:compar_snap}, the rim formed at the edge of the crown is
accurately captured in the simulation, but there is small shift at the base of the sheet.  We attribute the latter to various perturbations (e.g. surface waves generated when the drop detaches from the needle or air resistance as the drop falls) that cause the drop in the experimental system to deviate from the spherical geometry used in simulations  (see for instance \citet{Thoraval:2013a}).

\subsection{Numerical results}
 The aim of our numerics was to find a phenomenological law for the evolution of the characteristics of the jet as a prelude to developing a reduced model of the rim's dynamics.  In particular, we measured the size of the jet when it first emerges from neck $e_o$, the time-dependence of the rim radius $r$, and the vertical $H_c$ and horizontal $R_c$ distance of the rim from the impact center 

\begin{figure}
\begin{center}
         \includegraphics[width=\textwidth]{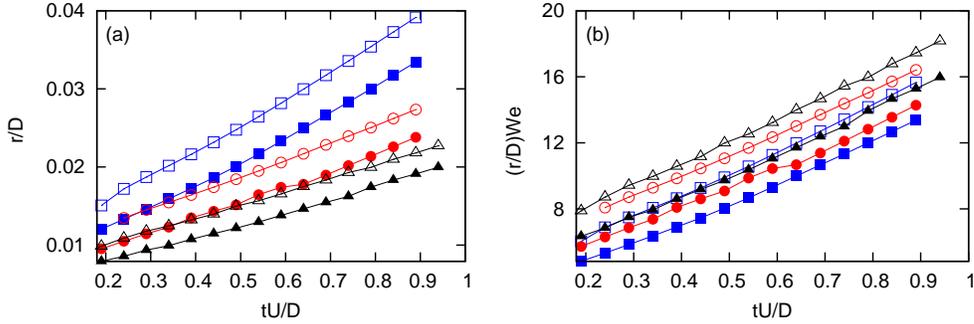}
\end{center}
\caption{Scaling of rim radius evolution. Open symbols correspond to \Rey=500  and filled symbols to \Rey=1000.
(a) Rim radius versus time for \Web=400 ($\square$), \Web=600 ($\bigcirc$), 
and \Web=800 ($\triangle$).
Note the general trend of decreasing rim size with increasing \Rey\  and \Web. (b) Data from (a) replotted with the rim radius rescaled to $\Web\  r/D$.}
\label{fig:simul_rim}
\end{figure}

\begin{figure}
\begin{center}
      \includegraphics[width=\textwidth]{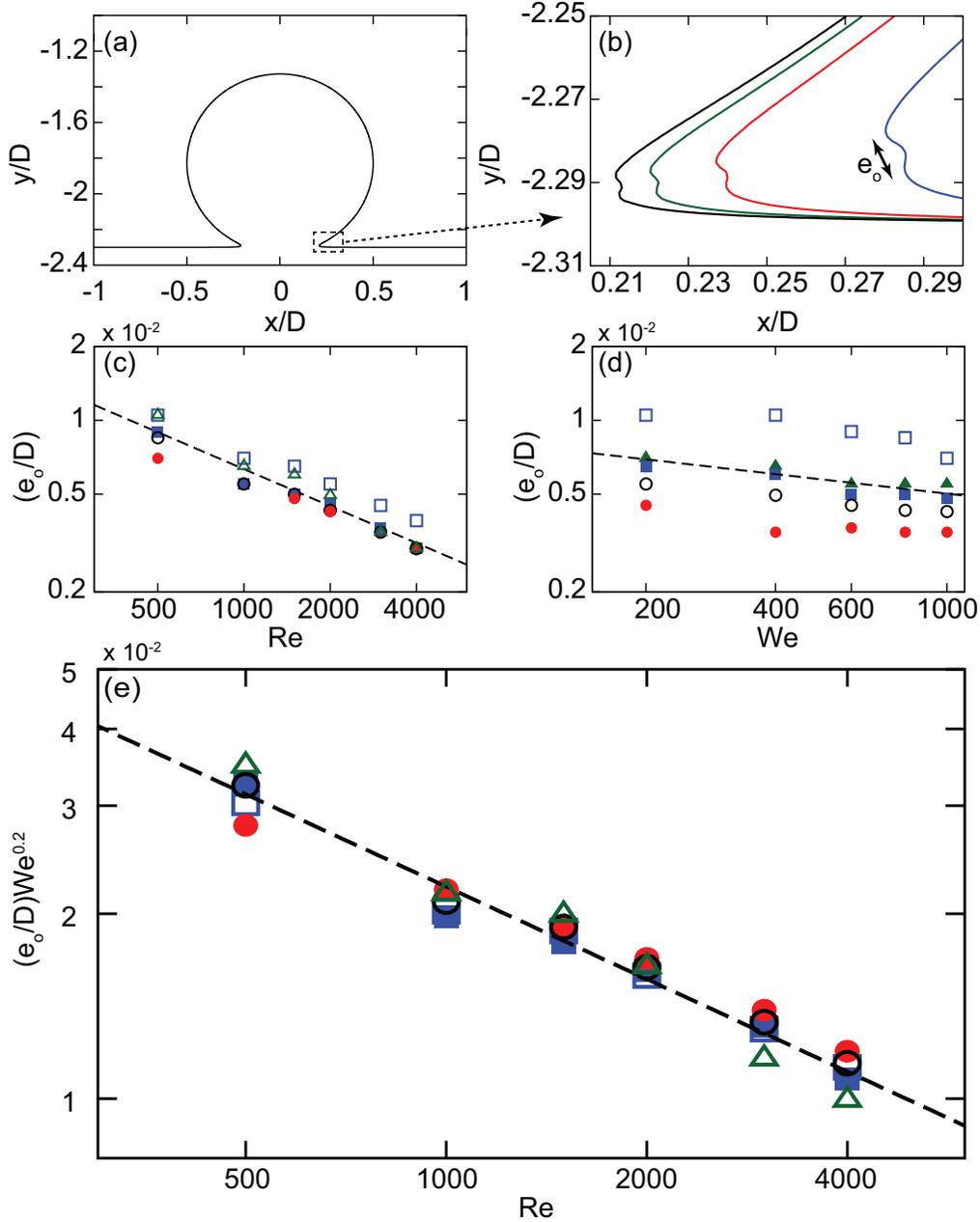}
\end{center}
\caption{Emergence of jet. (a) Free surface profile from simulations just 
after impact.  (b) Magnified view of neck between the drop and the pool (see dashed box in (a)) at the moment the jet emerges from the 
neck for $\Web=400$ and $\Rey= 2000, 1500, 1000, 500$ (from left to right). The arrow indicates the thickness of the jet as it emerges 
from the neck $e_o$.  Note that the jet emerges at a distance from the center that decreases with increasing \Rey. (c) $e_o$ versus 
\Rey\ for  \Web~= 200 (red $\square$), 400 (green $\triangle$), 600 (blue $\blacksquare$), 800 (red $\circ$), 1000 (red $\bullet$).  The dashed 
line is the power law $\Rey^{-0.5}$.  (d) $e_o$ versus \Web~for \Rey~= 500 (blue $\square$), 1000 (green $\triangle$), 
1500 (blue $\blacksquare$), 2000 (black $\circ$), 3000 (red $\bullet$). The dashed line is the power law $\Web^{-0.2}$. 
(e) Scaled  jet thickness versus \Rey.  The color scheme is the same as in (c).  The dashed line is the power law $0.7 \Rey^{-0.5}$.}
\label{fig:init_snap2}
\end{figure}

 The time evolution of the rim radius is shown in Fig.~\ref{fig:simul_rim}(a) for six different parameter sets. We varied \Rey\ and \Web\ (by changing only the liquid properties: $\mu_L$ and $\sigma_L$), keeping ${h}=0.2$ and the impact speed ${U}$ constant. These data show that $r$ grows linearly in time and
 the overall scale decreases with increasing \Rey\ and \Web. As shown in \ref{fig:simul_rim}(b), the radius depends linearly in time; 
moreover, the slope scales as \Web$^{-1}$:
  
\begin{equation}
r=\frac12 e_o +\frac{a}{\Web}Ut
\label{eq:rim_scaling_model}
\end{equation}
where $a$ is a real constant and $e_o$ is the initial diameter of the rim as it emerges from the neck(see Fig.~\ref{fig:init_snap2}).  Note that according to Eq.~\ref{eq:rim_scaling_model} the time at which the jet emerges  and $t=0$ are identical; the difference in these times in our numerical calculations is no more than 5~$\mu$s and thus we assume it is negligible.  

We measured the size of the jet when it first emerges from the neck connecting the drop and the pool from the distance between the inflection points on the neck as shown in Fig.~\ref{fig:init_snap2}(b).  The data for thirty different combinations of \Web\ (200, 400, 600, 800, 1000) and \Rey\ (500, 1000, 1500, 2000, 3000, 4000)
are shown in Fig.~\ref{fig:init_snap2}(b-d).  We find empirically that 
\begin{equation}
e_o \propto D\Web^{-0.2}\Rey^{-0.5}
\end{equation}
Furthermore, the distance of the neck from the center of the drop when the jet first emerges, usually called \emph{contact length}, decreases with increasing \Rey\ (see Fig. \ref{fig:init_snap2}(b); the center of the drop is at x/D=0) in agreement with the results of \citet{Coppola11}.  Combining the scaling for $e_o$ and $r$ yields: 
 \begin{equation}
  \frac{r}{D} \simeq \frac{a}{\Web} \frac{Ut}{D}+\frac{b}{\Web^{0.2}\Rey^{0.5}}
 \label{eq:radius}
 \end{equation}
 where $a$ and $b$ are constants with values obtained from fits to the data of 13 and 0.35 respectively.

\begin{figure}
\begin{center}

          \includegraphics[width=\textwidth]{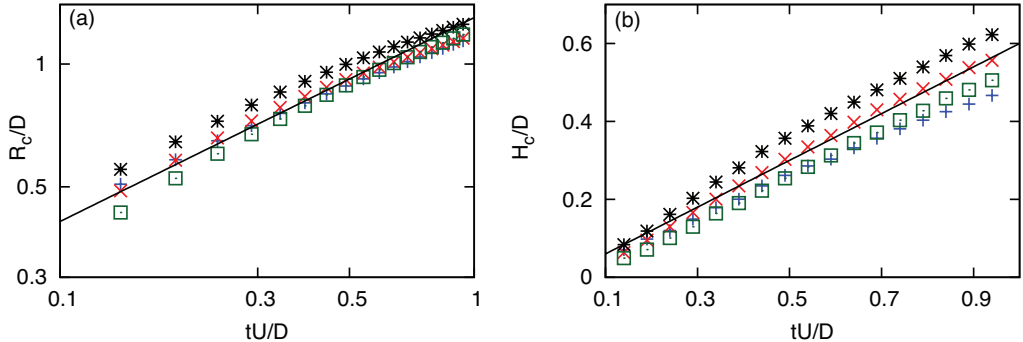}

\end{center}
\caption{Rim characteristics for  \Web=600 and \Rey=1000 (black $\ast$), \Web=400 and \Rey=500 (blue $+$), \Web=400 and \Rey=1000 (red $\times$) and \Web=600 and \Rey=500 (green $\square$).
 (a) Crown radius versus time. The line is the power law $(tU/D)^{0.5}$. 
(b) Crown height versus time. The linear line is $y=0.6(tU/D)$.
}
\label{fig:D}
\end{figure}

We measured the radius of the crown $R_c$ from the horizontal distance of the leading edge of the rim to the impact center as a function of time and the vertical distance of the crown above the undisturbed pool $H_c$.  These data are plotted
in Fig.~\ref{fig:D} for various combinations of \Rey\ and \Web; $R_c\sim t^{1/2}$  and $H_c\sim t$, largely independent of \Rey\ and \Web.
(\citet{Liang13} observed the same behavior.)  Hence the acceleration is primarily along the radial direction.

\section{Modeling wavelength selection of the rim instability}
\citet{Zhang:2010} showed good agreement of experiments with a stability analysis based on the Rayleigh-Plateau instability that ignored 
the connection of the rim to the jet and the deceleration of the rim.  \citet{fullana99} however argued that the former effect is 
essential because it saturates the growth of the Rayleigh-Plateau instability and thus prevents the formation of secondary droplets 
within the relevant timescale.  Moreover, \citet{Krechetnikov09a} also argued that acceleration is important because their measurements, 
albeit in a different regime from the crown splash, were consistent with a Ritchmyer-Meshkov instability. Below we examine the effects 
of the connection of the jet to the rim and the deceleration of the rim with a model that merges the long-wavelength linear stability 
analysis of \citet{Agbaglah13} and the results of our numerical calculations.

 \citet{Agbaglah13} analyzed in the inviscid limit the stability of the rim of a flat sheet-like jet of constant thickness and speed. 
 Importantly, it incorporate both acceleration and capillarity. Their model agrees well with
 full numerical simulations of that scenario.  The crown sheet however is curved, its thickness and speed are nonuniform, and its 
geometry is cylindrical.  We gloss over these difficulties in an attempt to construct a minimal model of the splash. 
The model of \citet{Agbaglah13} requires as inputs the ratio of the jet thickness to rim radius $e/r$ and the acceleration of the rim.
 We supply the former from the scaling relationship for $e_0$ under the assumption that this value is representative of $e$ and the latter from the rim position. 
Based on this composite model we predict the most unstable wavelength of a crown 
splash for a wide range of \Web\ and \Rey\ numbers.  In particular, for $\Rey=760$ and $\Web=1060$  for which there is high quality 
experimental data available we find excellent agreement.

\subsection{Wavelength selection and secondary droplet production}
\begin{figure}
\begin{center}

          \includegraphics[width=\textwidth]{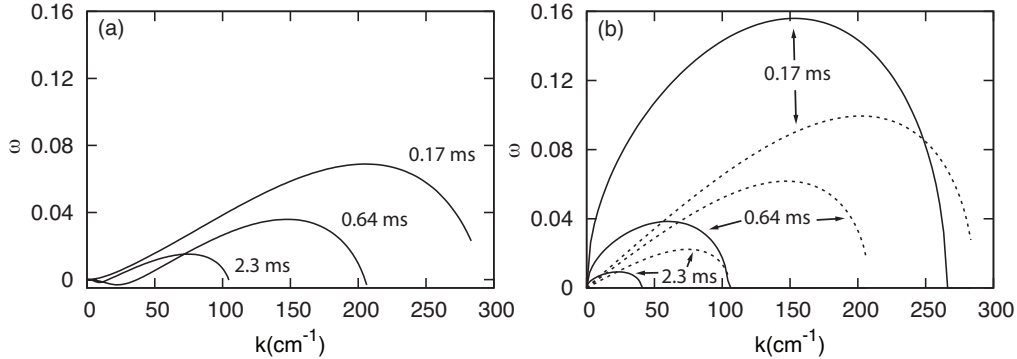}

\end{center}
\caption{Dispersion curves for \Web=760 and \Rey=1060. Growth rate vs wavenumber at various times after impact for (a) the theory of \citet{Agbaglah13} and (b) inviscid Rayleigh-Plateau (\dashed{0.5mm}{0.5mm}) and Rayleigh-Taylor (\solid{2.5mm})). }

\label{fig:omega}
\end{figure}

Using the rim radius and acceleration obtained in the previous sections as inputs for the linear stability calculation of \citep{Agbaglah13}, we computed the growth rate $\omega$ as a function of wavenumber $k$, and from this dispersion relationship the power and wavelength of the most unstable mode.  The dispersion relationship for \Web=760 and \Rey=1060 is plotted in Fig.~\ref{fig:omega}(a).  Figure~\ref{fig:omega}(b) shows the equivalent dispersion curves for the classical Rayleigh-Plateau and Rayleigh-Taylor instabilities.  The most unstable wavelength of the classical Rayleigh-Plateau most closely tracks the most unstable wavelength of our stability calculation.

\begin{figure}
\begin{center}

    \includegraphics[width=\textwidth]{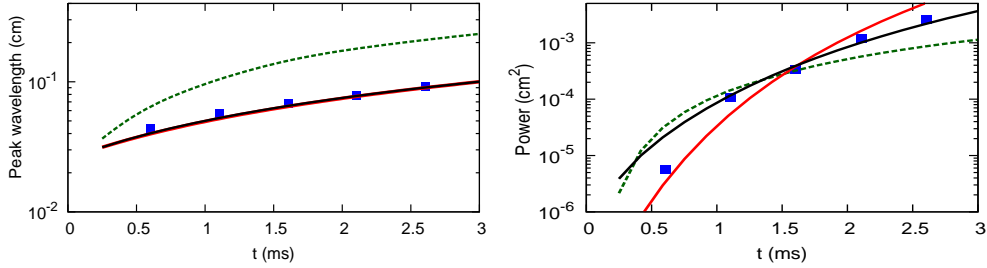}

\end{center}

\caption{Most unstable mode. (Left) Peak wavelength and (right) power in peak wavelength versus time for \Web=760 and \Rey=1060 
from experiments (blue $\blacksquare$)~\citep{Zhang:2011}, Rayleigh-Taylor theory with $\Psi_0/D=1.9\times10^{-3}$ (dashed green line), Rayleigh-Plateau model with $\Psi_0/D=0.3\times10^{-3}$ (red line), the theory of \citep{Agbaglah13} with $\Psi_0/D=7.5\times10^{-3}$ (black line).  }
\label{fig:wave}
\end{figure}

From the growth rate we compute the amplitude of the $n$th mode:
\begin{equation}
\Psi_n(t)=\Psi_o \max\limits_k\left(\exp\left\{\int_{0}^t \omega\left(k=\frac{n}{R_c},t'\right) dt'\right\}\right)
\end{equation}
where $\Psi_o$ is taken as a constant independent of wavelength under the assumption that the initial microscopic corrugations present 
in the system are distributed like broadband white noise.  The peak mode is selected as the mode with the maximum amplitude at any 
given time.  The peak wavelength and its amplitude are compared with the experimental measurements of \citet{Zhang:2010} 
in Fig.~\ref{fig:wave}.  In addition, we plot the classical inviscid Rayleigh-Plateau and Rayleigh-Taylor theory~\citep{Chandrasekhar81}
in Fig.~\ref{fig:wave} for comparison. In all three calculations $\Psi_o$ is taken as an adjustable parameter chosen to maximize the 
correspondence between the calculation and the experimental measurements.  These data show that our model is in excellent agreement 
with measurements. 

\begin{figure}
       \centerline{ \includegraphics[width=\textwidth]{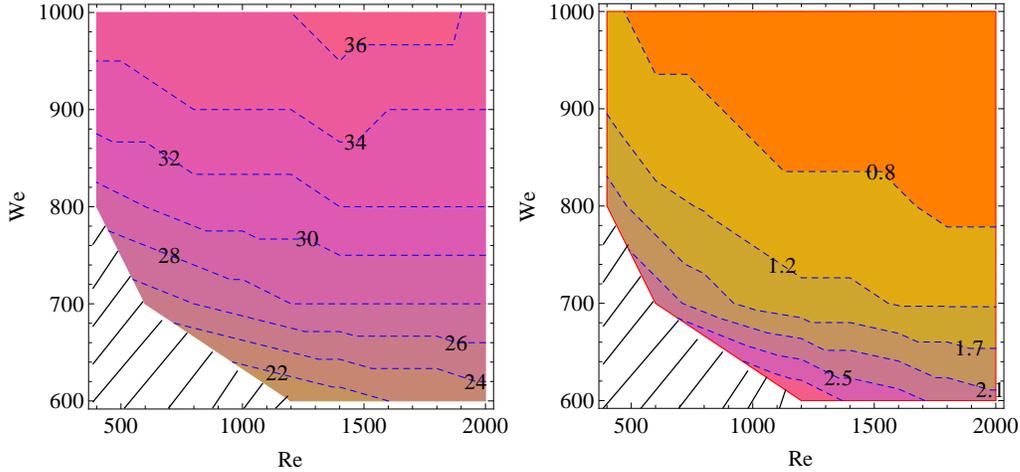}}
\caption{Model predictions. \emph{Left:} Number of secondary droplets  as a function of \Rey\ and \Web.  \emph{Right:} Time (in ms)
when amplitude of the most unstable mode becomes comparable to the diameter of the rim as a function of \Rey\ and \Web.  We take this time as a threshold for the nonlinear regime and the initiation of the droplet pinchoff process.  Note that while the power depends on $\Psi_o$, the peak wavelength does not.  The prediction of the latter is arrived at without adjustable parameters.}
\label{fig:droplet}
\end{figure}

We can predict the number of secondary droplets and the timescale for their onset over the entire parameter range of the crown splash with a few further assumptions.  We repeated the above calculation of $\Psi_n$ on a regular grid of parameter values (\Rey, \Web)
assuming that $\Psi_o$ is the same for all parameters and equal to the value obtained from the comparison in figure~\ref{fig:wave}
(right).  We further assume that when the amplitude of the peak corrugation grows to a size comparable to the diameter of the rim 
(i.e. $\Psi_n\backsimeq 2r$, the choice of $2r$ is motivated by the fact that the antisymmetric part of the perturbation is dominant in the nonlinear regime (e.g. see Fig. 9 and 10 
of \citet{Agbaglah13})) nonlinear effects become dominant and lock in the most unstable mode.  Lastly, we 
assume that the number of droplets is given by the peak mode $n^{*}$ as suggested by the experiments of \citet{Zhang:2010}.
Figure~\ref{fig:droplet} shows the resulting predictions.  These data predict that the number of secondary droplets increase with 
both \Rey\ and \Web.
No secondary droplets are predicted for the hashed region of Fig.~\ref{fig:droplet}, since $\Psi_n < 2r$ for the entire timescale of splashing (i.e. $\tilde{t}\lesssim 2.7$).

\begin{figure}
\centerline{\includegraphics[width=\textwidth]{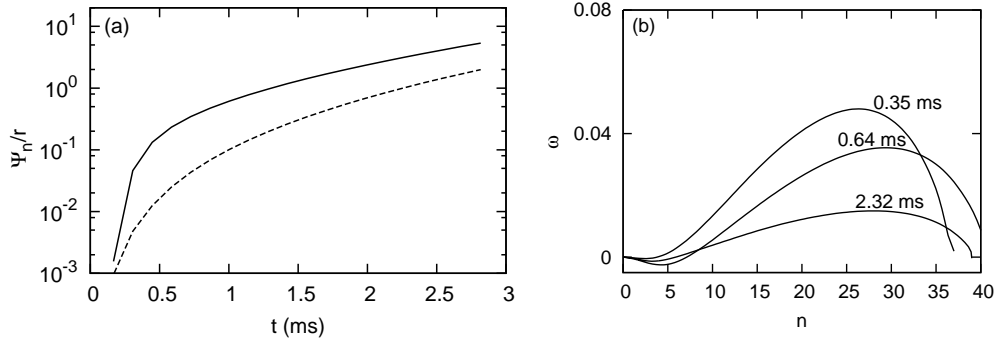}}
\caption{(a) Amplitude versus time for \Web=760 and \Rey=1060 with acceleration (solid line) and without the acceleration (dashed line) for the same time 
evolution of the rim radius and the same initial amplitude $\Psi_o$. (b) Same dispersion curve as in Fig.~\ref{fig:omega}(a) plotted 
versus mode number.}
\label{fig:ampli}
\end{figure}

\section{Discussion and Conclusion}
We performed axisymmetric simulations of drop impact on a thin film in the crown splash regime using a volume-of-fluid implementation of the Navier-Stokes equations and validated these by comparing with experimental profiles obtained using high speed X-ray imaging.  By combining the result of our axisymmetric simulations with the theory of \citet{Agbaglah13} for symmetry breaking of a flat sheet, we predict the most unstable mode of the axial symmetry breaking instability in a crown splash.  With the further assumption, supported by experiments, that the crests of the most unstable mode are the origin of secondary droplets, we predict the secondary droplet production throughout the crown splash regime ($\Web<1000$, $\Rey<2000$).

From simulations we obtain scaling relations for the position ($R_c$ and $H_c$) and the size ($r$) of the rim, and the initial thickness ($e_o$) of the jet. These are then used in the linear stability theory of \citet{Agbaglah13} to obtain the most  unstable wavelength and hence the number of secondary droplets.  No adjustable parameters are used to predict the most unstable wavelength; a single adjustable parameter, the initial amplitude of the corrugation $\Psi_o$ that initiates the instability, is used to predict the magnitude of the instability.  Our predictions are in excellent agreement with the experimental measurements of \citet{Zhang:2010}. 

The agreement between experiments and theory could be perhaps improved if we allowed for a dependence of $\Psi_o$ on \Rey\ and \Web.  However, in absence of experimental evidence that would warrant this more complicated scenario, we chose $\Psi_o$ to be independent of wavelength, \Rey~and \Web.

The debate about the cause of splashing has largely revolved around the relative importance of capillarity versus acceleration, or -- as it is frequently phrased in the literature -- whether the instability is primarily Rayleigh-Plateau-like or Rayleigh-Taylor-like.  With our simulation results we can quantitatively examine this question in the crown splash regime.  As shown by Fig.~\ref{fig:omega}, our results favor a Rayleigh-Plateau-like instability as the primary mechanism for wavelength selection.  

Our calculation however, differs from the classical Rayleigh-Plateau in that it includes the effects of deceleration of the rim and the connection of the rim to the jet.  The combination of these two effects yields a better fit to the experimental data.  The rim-jet connection weakens the destabilizing role of capillarity.  For example, consider the extreme case that prevails when the jet first forms: the jet ends in a semicircle with a diameter equal to the sheet thickness.  Unlike the case of a full cylinder, there is no  surface energy gain for transverse corrugation for the free end of the jet.  Thus, we expect that the classical Rayleigh-Plateau theory which is for a full cylinder will produce a greater growth rate than our theory; this is indeed the case as shown by Fig.~\ref{fig:omega}.  The deceleration of the rim however, counteracts the dampening effect of the rim-jet connection.  As shown in Fig.~\ref{fig:ampli}(a), the amplitude of the most unstable mode is significantly more amplified when acceleration is included.    \citet{roisman10} reached a similar conclusion in the context of flat sheets.   

How can we reconcile the obvious effect of acceleration on the magnitude of the mode with its absence in wavelength selection?  The deceleration phase is too short-lived to effect wavelength selection.  As the rim approaches the Taylor-Culick limiting speed (relative to the jet speed), acceleration and its destabilizing effect vanish.   We find in our simulations that this happens rapidly.  The legacy of this initial acceleration phase is a greater initial amplitude for the subsequent evolution of the rim due to the capillary instability.  Moreover, since the Rayleigh-Plateau and Rayleigh-Taylor instabilities have similar peak wavelengths because they both originate from the same driving force, the principal contribution of the acceleration instability is centered around the modes susceptible to the capillary instability.  

\section{Acknowledgements}
The authors thank the James S. McDonnell Foundation for support through a 21st Century
Science Initiative in Studying Complex Systems Research Award; L. V. Zhang, P. Ray and C. Josserand for valuable discussions;
 and S. Veerapaneni, S. Zaleski and the \emph{Institut~Jean~le~Rond d'Alembert} for sharing their computational resources.

 \bibliographystyle{jfm}
%
%
%
 \bibliography{splash}

\begin{thebibliography}{27}
\expandafter\ifx\csname natexlab\endcsname\relax\def\natexlab#1{#1}\fi

\bibitem[Agbaglah {\em et~al.\/}(2011)Agbaglah, Delaux, Fuster, Hoepffner,
  Josserand, Popinet, Scardovelli \& Zaleski]{Gerris-CRAS}
{\sc Agbaglah, G., Delaux, S., Fuster, D., Hoepffner, J., Josserand, C.,
  Popinet, S.~Ray, P., Scardovelli, R. \& Zaleski, S.} 2011 Parallel simulation
  of multiphase flows using octree adaptivity and the volume-of-fluid method.
  {\em C. R. Mecanique.\/} {\bf 339}, 194--207.

\bibitem[Agbaglah {\em et~al.\/}(2013)Agbaglah, Josserand \&
  Zaleski]{Agbaglah13}
{\sc Agbaglah, G., Josserand, C. \& Zaleski, S.} 2013 Longitudinal instability
  of a liquid rim. {\em Phys. Fluids\/} {\bf 25}, 022103.

\bibitem[Bremond \& Villermaux(2006)]{Bremond06}
{\sc Bremond, N. \& Villermaux, E.} 2006 Atomization by jet impact. {\em
  Journal of Fluid Mechanics\/} {\bf 549}, 273--306.

\bibitem[Chandrasekhar(1981)]{Chandrasekhar81}
{\sc Chandrasekhar, S.} 1981 Hydrodynamic and hydromagnetic stability. {\em
  Dover, New York\/} .

\bibitem[Coppola {\em et~al.\/}(2011)Coppola, Rocco \& de~Luca]{Coppola11}
{\sc Coppola, G., Rocco, G. \& de~Luca, L.} 2011 Insights on impact of a plane
  drop on a thin liqui film. {\em Phys. Fluids.\/} {\bf 23}, 022105.

\bibitem[Cossali {\em et~al.\/}(1997)Cossali, Coghe \& Marengo]{cossali97}
{\sc Cossali, G.~E., Coghe, A. \& Marengo, M.} 1997 Impact of a single drop on
  a wetted solid surface. {\em Experiments in Fluids\/} {\bf 22}~(6), 463--472.

\bibitem[Deegan {\em et~al.\/}(2008)Deegan, Brunet \& Eggers]{deegan08}
{\sc Deegan, R.~D., Brunet, P. \& Eggers, J.} 2008 Complexities of splashing.
  {\em Nonlinearity\/} {\bf 21}~(1), C1--C11, 0951-7715.

\bibitem[Fullana \& Zaleski(1999)]{fullana99}
{\sc Fullana, J.~M. \& Zaleski, S.} 1999 Stability of a growing end rim in a
  liquid sheet of uniform thickness. {\em Physics of Fluids\/} {\bf 11}~(5),
  952--954.

\bibitem[Fuster {\em et~al.\/}(2009)Fuster, Agbaglah, Josserand, Popinet \&
  Zaleski]{fuster:2009a}
{\sc Fuster, Daniel, Agbaglah, Gilou, Josserand, Christophe, Popinet, Stephane
  \& Zaleski, Stephane} 2009 Numerical simulation of droplets, bubbles and
  waves: state of the art. {\em Fluid Dynamics Research\/} {\bf 41}~(6),
  065001.

\bibitem[Gueyffier \& Zaleski(1998)]{Gueyffier98}
{\sc Gueyffier, D. \& Zaleski, S.} 1998 Finger formation during droplet impact
  on a liquid film. {\em C.R. Acad. Sci., Ser. IIb: Mec., Phys., Chim.,
  Astron\/} {\bf 326}~(12), 839--844.

\bibitem[Krechetnikov(2010)]{Krechetnikov10}
{\sc Krechetnikov, R.} 2010 Stability of liquid sheet edges. {\em Physics of
  Fluids\/} {\bf 22}, 092101.

\bibitem[Krechetnikov \& Homsy(2009)]{Krechetnikov09a}
{\sc Krechetnikov, Rouslan \& Homsy, George~M.} 2009 Crown-forming instability
  phenomena in the drop splash problem. {\em J Colloid Interface Sci\/} {\bf
  331}~(2), 555--9.

\bibitem[Liang {\em et~al.\/}(2013)Liang, Guo, Shen \& Yang]{Liang13}
{\sc Liang, G., Guo, Y., Shen, S. \& Yang, Y.} 2013 Crown behavior and bubble
  entrainment during a drop impact on a liquid film. {\em Theor. Comput. Fluid
  Dyn.\/} {\bf DOI: 10.1007/s00162-013-0308-z}.

\bibitem[Popinet(2003{\natexlab{{\em a\/}}})]{Pop03}
{\sc Popinet, S.} 2003{\natexlab{{\em a\/}}} Gerris: a tree-based adaptive
  solver for the incompressible euler equations in complex geometries. {\em J.
  Comp. Phys.\/} {\bf 190}, 572--600.

\bibitem[Popinet(2003{\natexlab{{\em b\/}}})]{Gerris}
{\sc Popinet, S.} 2003{\natexlab{{\em b\/}}} Gerris flow solver.
  http://gfs.sourceforge.net/.

\bibitem[Popinet(2009)]{GerrisVOF}
{\sc Popinet, S.} 2009 An accurate adaptive solver for surface-tension-driven
  interfacial flows. {\em J. Comput. Phys.\/} {\bf 228}, 5838--5866.

\bibitem[Rieber \& Frohn(1999)]{Rieber99}
{\sc Rieber, M. \& Frohn, A.} 1999 A numerical study on the mechanism of
  splashing. {\em Int. J. Heat Fluid Flow\/} {\bf 20}~(5), 455--461.

\bibitem[Rioboo {\em et~al.\/}(2003)Rioboo, Bauthier, Conti, Voue \&
  De~Coninck]{rioboo03}
{\sc Rioboo, R., Bauthier, C., Conti, J., Voue, M. \& De~Coninck, J.} 2003
  Experimental investigation of splash and crown formation during single drop
  impact on wetted surfaces. {\em Experiments in Fluids\/} {\bf 35}~(6),
  648--652.

\bibitem[Roisman(2010)]{roisman10}
{\sc Roisman, I.~V.} 2010 On the instability of a free viscous rim. {\em
  Journal of Fluid Mechanics\/} {\bf 661}, 206--228.

\bibitem[Roisman {\em et~al.\/}(2007)Roisman, Gambaryan-Roisman, Kyriopoulos,
  Stephan \& Tropea]{roisman07}
{\sc Roisman, I.~V., Gambaryan-Roisman, T., Kyriopoulos, O., Stephan, P. \&
  Tropea, C.} 2007 Breakup and atomization of a stretching crown. {\em Physical
  Review E\/} {\bf 76}~(2).

\bibitem[Roisman {\em et~al.\/}(2006)Roisman, Horvat \& Tropea]{roisman06}
{\sc Roisman, I.~V., Horvat, K. \& Tropea, C.} 2006 Spray impact: Rim
  transverse instability initiating fingering and splash, and description of a
  secondary spray. {\em Physics of Fluids\/} {\bf 18}~(10).

\bibitem[Thoraval {\em et~al.\/}(2012)Thoraval, Takehara, Etoh, Popinet, Ray,
  Josserand, Zaleski \& Thoroddsen]{Thoraval:2012a}
{\sc Thoraval, Marie-Jean, Takehara, Kohsei, Etoh, Takeharu~Goji, Popinet,
  Stephane, Ray, Pascal, Josserand, Christophe, Zaleski, Stephane \&
  Thoroddsen, Sigurdur~T.} 2012 von karman vortex street within an impacting
  drop. {\em Physical Review Letters\/} {\bf 108}~(26).

\bibitem[Thoraval {\em et~al.\/}(2013)Thoraval, Takehara, Etoh \&
  Thoroddsen]{Thoraval:2013a}
{\sc Thoraval, M.~J., Takehara, K., Etoh, T.~G. \& Thoroddsen, S.~T.} 2013 Drop
  impact entrapment of bubble rings. {\em Journal of Fluid Mechanics\/} {\bf
  724}, 234--258.

\bibitem[Worthington(1879)]{Worthington1879}
{\sc Worthington, A.~M.} 1879 On the spontaneous segmentation of a liquid
  annulus. {\em Proc. Phys. Soc. London\/} {\bf 30}, 49--60.

\bibitem[Yarin \& Weiss(1995)]{Yarin95}
{\sc Yarin, A.~L. \& Weiss, D.~A.} 1995 Impact of drops on solid-surfaces -
  self-similar capillary waves, and splashing as a new-type of kinematic
  discontinuity. {\em Journal of Fluid Mechanics\/} {\bf 283}, 141--173.

\bibitem[Zhang {\em et~al.\/}(2010)Zhang, Brunet, Eggers \& Deegan]{Zhang:2010}
{\sc Zhang, L.~V., Brunet, P., Eggers, J. \& Deegan, R.~D.} 2010 Wavelength
  selection in the crown splash. {\em Physics of Fluids\/} {\bf 22}~(12).

\bibitem[Zhang {\em et~al.\/}(2011)Zhang, Toole, Fezzaa \& Deegan]{Zhang:2011}
{\sc Zhang, L.~V., Toole, J., Fezzaa, K. \& Deegan, R.~D.} 2011 Evolution of
  the ejecta sheet from the impact of a drop with a deep pool. {\em J. Fluid
  Mech.\/} {\bf 396}, 1--11.

\end{thebibliography}

\end{document}